\documentclass[12pt]{iopart}
\usepackage{graphicx}

\begin{document}

\title{
Non-metallic, non-Fermi-liquid resistivity of FeCrAs from 0 to 17 GPa
}

\author{F F Tafti$^{1,2}$, W Wu$^1$ and S R Julian$^{1,3}$}

\address{1. Department of Physics, University of Toronto, 60 St.\ George Street, 
            Toronto, ON, M5S 1A7, Canada}
\address{2. Physics Department, University of Sherbrooke, 2500 Boul.\ Universit\'e, 
            Sherbrooke, QC, J1K 2R1, Canada}
\address{3. Canadian Institute for Advanced Research, Quantum Materials Program,
            180 Dundas St.\ W., Suite 1400, Toronto, ON, M5G 1Z8, Canada}
\ead{fazel.fallah.tafti@usherbrooke.ca} 

\begin{abstract} 
An unsual, non-metallic resistivity of the 111 iron-pnictide 
  compound FeCrAs is shown to be relatively unchanged under pressures of up to 17 GPa. 
Combined with our previous finding that this non-metallic behaviour 
  persists from at least 80 mK to 800 K, 
  this shows that the non-metallic phase is exceptionally robust. 
Antiferromagnetic order, with a N\'eel temperature $T_N \sim 125$ K 
  at ambient pressure, is suppressed by pressure at a rate of 
  $7.3 \pm 0.1$ K/GPa, falling to below 50 K at 10 GPa. 
We conclude that 
  formation of a spin-density wave gap at $T_N$ does not 
  play an important role in the 
  non-metallic resistivity of FeCrAs at low temperatures.  
\end{abstract}

\pacs{71.10.Hf,71.27.+a,75.15.Rn}
\maketitle

\section{Introduction:}
\label{sec:intro}

According to band theory, 
  the resistivity of an insulator rises exponentially with decreasing temperature 
  because thermally activated carriers are being frozen out.  
The resistivity of a metal, in contrast, falls with decreasing temperature, 
  because the density of carriers is fixed while thermally excited 
  scattering is frozen out. 
However there are exceptions -- `non-metallic metals' -- 
  where the resistivity 
  rises at low temperature as the temperature falls, but without 
  a gap opening in the electronic excitation spectrum. 
Some of these exceptions are well understood, for example Kondo systems or 
  disordered metals showing weak or strong localization according to whether 
  they have high or low carrier densities. 
There are also exceptions that are not understood, such as the 
  underdoped cuprates \cite{ando95,dobrosavljevic12}, and some 
  disordered heavy fermion compounds \cite{maple95}. 
These systems may have high carrier density yet display 
  strongly non-metallic resistivity as the temperature is varied between room 
  temperature and $T\rightarrow 0$ K. 
Moreover, in the $T\rightarrow 0$ K limit the non-metallic heavy fermion systems show 
  non-Fermi-liquid properties in their 
  resistivity, linear specific heat coefficient, and magnetic susceptibility \cite{maple95}:
  $\rho \propto T$, $C/T \propto -\ln(T)$, $\chi(T) \propto 1 - c\sqrt{T}$. 

Recently, the high temperature resistivity of some iron-pnictides 
  has also been found to be flat, or slowly rising, with decreasing 
  temperature, even in undoped parent compounds where disorder levels 
  are low \cite{stewart11}.  
In these systems, however, the resistivity generally begins to fall, 
  and becomes metallic, when magnetic order sets in, typically around 
  100 K to 200 K. 
It has been suggested that many-body effects, 
  arising from orbital degeneracy \cite{haule08} or spin-fluctuations 
  \cite{dai09}, 
  may be responsible for the anomalous temperature dependence of their  
  high-temperature resistivity. 

\begin{figure}
\begin{center}
\includegraphics[width=10cm]{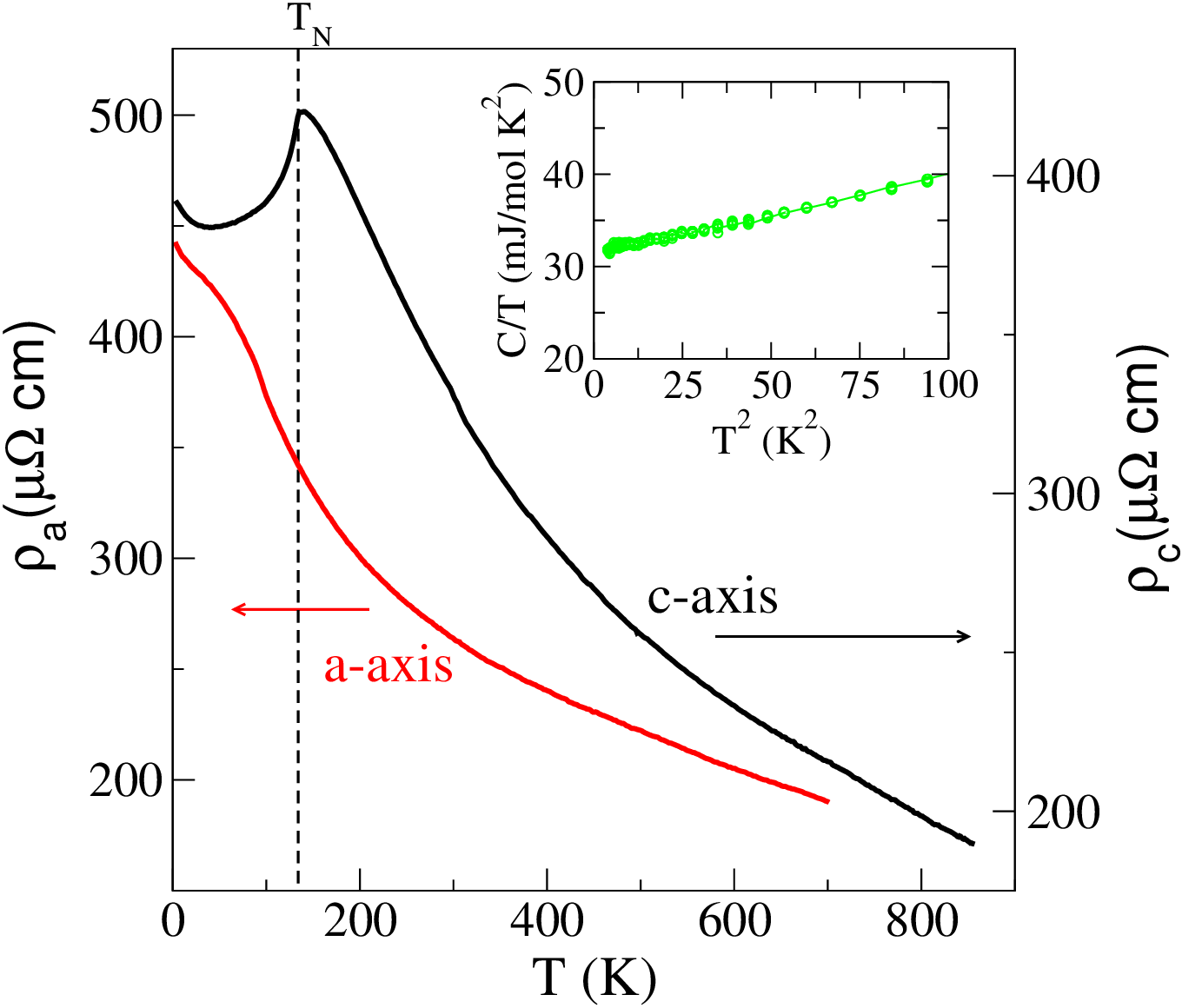}
\caption{Main figure: resistivity vs.\ temperature at ambient pressure 
\cite{wu09}. 
$\rho_{ab}$ rises monotonically with decreasing temperature, while 
the rise in $\rho_c$ is interrupted by a peak at the N\'eel temperature 
$T_N \sim 125$ K.  As $T\rightarrow 0$ K the resistivity has a negative 
slope with a sub-linear, non-Fermi-liquid power law.  The inset shows 
that, in contrast, the specific heat is Fermi-liquid-like at low temperature. 
The data are from reference \cite{wu09}. 
}
\label{fig:ambient}
\end{center}
\end{figure}

We recently showed that the hexagonal iron-pnictide, FeCrAs, is an 
  extreme example of such behaviour \cite{wu09}: 
  over four decades of temperature from above 800 K to 
  below 80 mK, single crystals show a resistivity in 
  the hexagonal plane that rises monotonically with decreasing temperature, 
  but without showing the presence of a gap (see figure \ref{fig:ambient}). 
The $c$-axis resistivity is similar, except for a local maximum 
  at a N\'eel ordering transition near 125 K. 
In the limit as $T\rightarrow 0$ K, the resisitivity shows a 
  non-metallic rise with a non-Fermi-liquid power law dependence on temperature. 
In contrast, the 
  specific heat, $C(T)$, and the ac-susceptibility, 
  $\chi(T)$, obey Fermi liquid power laws as $T \rightarrow 0$ K. 
This combination of non-Fermi-liquid transport with Fermi-liquid 
  thermodynamic properties is very unusual. 
Moreover, the linear coefficient of specific heat is large, 
  $C(T)/T \sim 30$ mJ/(mole\,K$^2$), as is the Pauli susceptibility, 
  showing that FeCrAs has a large 
  density of states at the Fermi energy.
If the carrier density were small,   
  as in some non-metallic metals, 
  this value of $C(T)/T$ would require extremely massive 
  quasiparticles, 
  which would itself be unusual for a $3d$-electron system.  
Band structure calculations, however, predict a large carrier density \cite{ishida96}, 
  with three large, three-dimensional Fermi surfaces \cite{sutton13}. 
According to neutron diffraction measurements, levels of disorder in 
  our single crystals are low \cite{wu11}.

The physics behind the non-metallic resistivity of FeCrAs is not known.  
Its general behaviour is somewhat reminiscent of systems that are on the border of 
  Anderson localization, in which the Fermi energy is close to the mobility edge 
  separating extended from localized states in disordered semiconductors \cite{mott70}. 
Such materials can show a non-metallic resistivity combined at low temperature with 
  a linear $C(T)$, 
  however the carrier density in FeCrAs seems much too high, and the level of disorder much too 
  low, for it to be in this regime. 
Thus, for example, in well-known cases with a non-metallic resistivity 
  such as phosphorus-doped silicon or non-stoichiometric Ce$_{3-x}$S$_4$ 
  \cite{thomas83,cutler69} the resistivity on the border of Anderson localization
  is over 100 times larger than we see in FeCrAs, 
  while in Si:P which has a linear specific heat at low temperature, $C(T)/T$ 
  is 300 times smaller than we see in FeCrAs \cite{paalanen88}.

An obstacle to understanding the physics of FeCrAs is the   
  antiferromagnetic transition near 125 K.  
As samples are cooled through this transition, 
  the $c$-axis resistivity falls abruptly in 
  the highest quality crystals, before continuing to rise again 
  at lower temperature, as shown in figure \ref{fig:ambient}.
This behaviour suggests that there is some spin-fluctuation scattering, 
  but it might also be compatible with an orbital mechanism. 
The $ab$-plane resistivity does not fall upon cooling below $T_N$, 
  indeed if anything the slope of $d\rho_{ab}/dT$ is steepest just below 
  $T_N$ (see figure \ref{fig:ambient}), 
  producing a weak `bump', or concave downwards region, below $T_N$. 

It would be interesting to know the low temperature limiting 
  behaviour of the resistivity in the absence of antiferromagnetic 
  order.  
A possible scenario is that the resistivity would saturate, 
  or even begin to fall, as $T\rightarrow 0$ K
  in the absence of magnetic order that opens a spin-density-wave gap over 
  part or all of the Fermi surface.
Or, at the other extreme, 
 perhaps a full gap would open over the entire Fermi surface, 
 leading to diverging resistivity as $T\rightarrow 0$ K, but 
 antiferromagnetism prevents the opening of this gap and leaves  
 a semi-metallic state with a small Fermi surface.  

The antiferromagnetic order in FeCrAs is itself unusual, 
  and indicative of some level 
  of frustration due to the P$\overline{6}$2m crystal structure, which 
  can be viewed as a triangular lattice of iron `trimers', plus a 
  distorted Kagome sublattice of Cr ions.  
Magnetic order is found only on the Cr sublattice, in the form of a  
  commensurate spin-density wave that triples the unit cell in the 
  hexagonal plane, while along the $c$-axis the moments are parallel 
  \cite{wu09,swainson10}.
The N\'eel temperature is low compared with the comparable tetragonal 
  systems Fe$_2$As and Cr$_2$As, which have $T_N \sim 350$ K \cite{katsuraki66} 
  and $T_N \sim 393$ K \cite{yuzuri60} respectively. 
In FeCrAs, the iron site is tetrahedrally 
  coordinated by As, as in the iron-pnictide superconductors, and 
  even below the antiferromagetic transition it does not display 
  a measurable magnetic moment in neutron scattering or 
  M\"ossbauer spectroscopy.
This is in agreement with band-structure calculations \cite{ishida96} 
  that found that the partial density of states on the iron sites is too low 
  to meet the Stoner criterion for magnetic moment formation, and indeed in a recent paper the 
  iron K$\beta$ x-ray emission spectrum from FeCrAs was used to provide a 
  non-magnetic reference \cite{gretarsson11}.
It should be noted, however, that in the related tetrahedral compound Fe$_2$As the iron 
  moment on the tetrahedrally coordinated site is 1.28 $\mu_B$ \cite{katsuraki66}, 
  suggesting that this site is close to the magnetic/non-magnetic moment-formation boundary, 
  and that frustration may also play a role in moment suppression on the iron site.  

The physics of frustrated metallic magnets still has many open questions \cite{lacroix10,julian12}. 
Based on the frustrated magnetic sublattices and the absence of a magnetic moment on the iron 
  sites, Rau et al.\  have put forward a theory that the anomalous behaviour of 
  FeCrAs arises from a `hidden spin liquid' on the 
  iron sublattice \cite{rau11}. 
In this picture, the conduction electrons fractionalize, and 
  anomalous transport is due to scattering of bosonic charge degrees of 
  freedom off of strong gauge fluctuations.

In this paper we try to determine whether the antiferromagnetism is 
  playing an important role, particularly in the $T \rightarrow 0$ K 
  limit, by using pressure to adjust $T_N$. 
We find that, despite suppressing $T_N$ by more than a factor of 
  two, and possibly all the way to 0 K in our highest pressure measurements, 
  the general behaviour of the 
  anomalous resistivity is not dramatically modified, 
  suggesting that the opening of a spin-density-wave gap does not play an 
  important role in the non-metallic resistivity
  of FeCrAs.

\section{Experiment:} \label{sec:expt}

We have carried out four-terminal resistivity measurements on single 
  crystal samples of FeCrAs at high pressure. 
Crystals were grown from a stoichiometric melt in an alumina 
  crucible within a sealed quartz tube.  
The material was melted twice, and then annealed at 900$^\circ$C 
  for 150 hours. 
Sample quality in FeCrAs is revealed by the sharpness of the resistive 
  transition at $T_N$, the value of $T_N$, and the temperature 
  at which glassy behaviour in the magnetic susceptibility 
  sets in. 
The crystals used in these measurements were from our highest 
  quality batch \cite{wu09}, in which $T_N \sim 125$ K in susceptibility measurements, 
  $T_N \sim 133$ K according to the cusp in the $c$-axis resistivity, 
  and in which glassy behaviour is very weak and only sets in below 10 K.  
Details of crystal growth and characterization can be found in 
  reference \cite{wu11}.

Electrical contacts to the samples were made with Dupont 6838 
  epoxy.  These had high resistivities at ambient pressure, 
  but under pressure they fell to the range of a few ohms. 

We pressurized two single crystals, one with $I\parallel c$ which 
  measures $\rho_c$, and the other with $I \perp c$, measuring $\rho_{ab}$. 
The $\rho_c$ sample had dimensions 
   $250\times 200\times 30\ \mu{\rm m}^3$. 
It was pressurized in a Moissanite anvil cell with 800 $\mu$m culets. 
  The gasket was beryllium-copper, with a 400 $\mu$m hole, insulated 
  with a mixture of alumina-powder and stycast 1266 epoxy.  
The $\rho_{ab}$ sample had dimensions $250\times100\times25\ \mu{\rm m}$ and 
  was pressurized in a diamond anvil cell 
  with 600 $\mu$m culets, using a fully hardened T301 stainless-steel gasket. 
Daphne oil 7373 was used at the pressure medium.
The pressure was determined at room temperature using ruby fluorescence; 
  the pressure may shift by up to $\sim 0.4$ GPa while the cell is cooled.  
The $\rho_c$ sample survived up to 10 GPa before an anvil broke, while 
  the $\rho_{ab}$ sample survived up to 17 GPa. 

In order to track the pressure dependence of $T_N$ we made use of 
  the peak in $\rho_c$ at $T_N$.
Unfortunately, $\rho_{ab}$  
  does not have a well-defined anomaly at $T_N$, as can be seen 
  in figure \ref{fig:ambient}, 
  so we could only follow $T_N$ vs.\ pressure with confidence up to 10 GPa. 

A possible concern with all high pressure measurements is 
  pressure-induced structural phase transitions. 
Among the 111 pnictides however, FeCrAs should be relatively immune to such transitions. 
The 111 pnictides come in three main crystal structures: tetrahedral, 
  hexagonal and orthorhombic, in order of decreasing unit cell volume 
  \cite{fruchart82}. 
Both Fe$_2$As and Cr$_2$As have the tetrahedral structure, and there is 
  only a narrow range of stability of the hexagonal phase around the 
  FeCrAs stoichiometry, thus FeCrAs must be just barely below the 
  volume criterion of stability for the hexagonal phase. 
We thus expect it to be able to withstand 
  quite a lot of compression before it transforms to the orthorhombic 
  phase, and indeed in our measurements we did not see any abrupt 
  changes in resistivity that would indicate a change of structure. 

Resistivity measurements were carried out at many pressures, as shown in figures 
  \ref{fig:rhocvsp} and \ref{fig:rhoabvsp}. At each pressure the temperature was 
  varied between room temperature and 2 K, using a dipping probe to control the 
  temperature. 

\section{Results:} \label{sec:results}

\begin{figure} 
\begin{center}
\includegraphics[width=10cm]{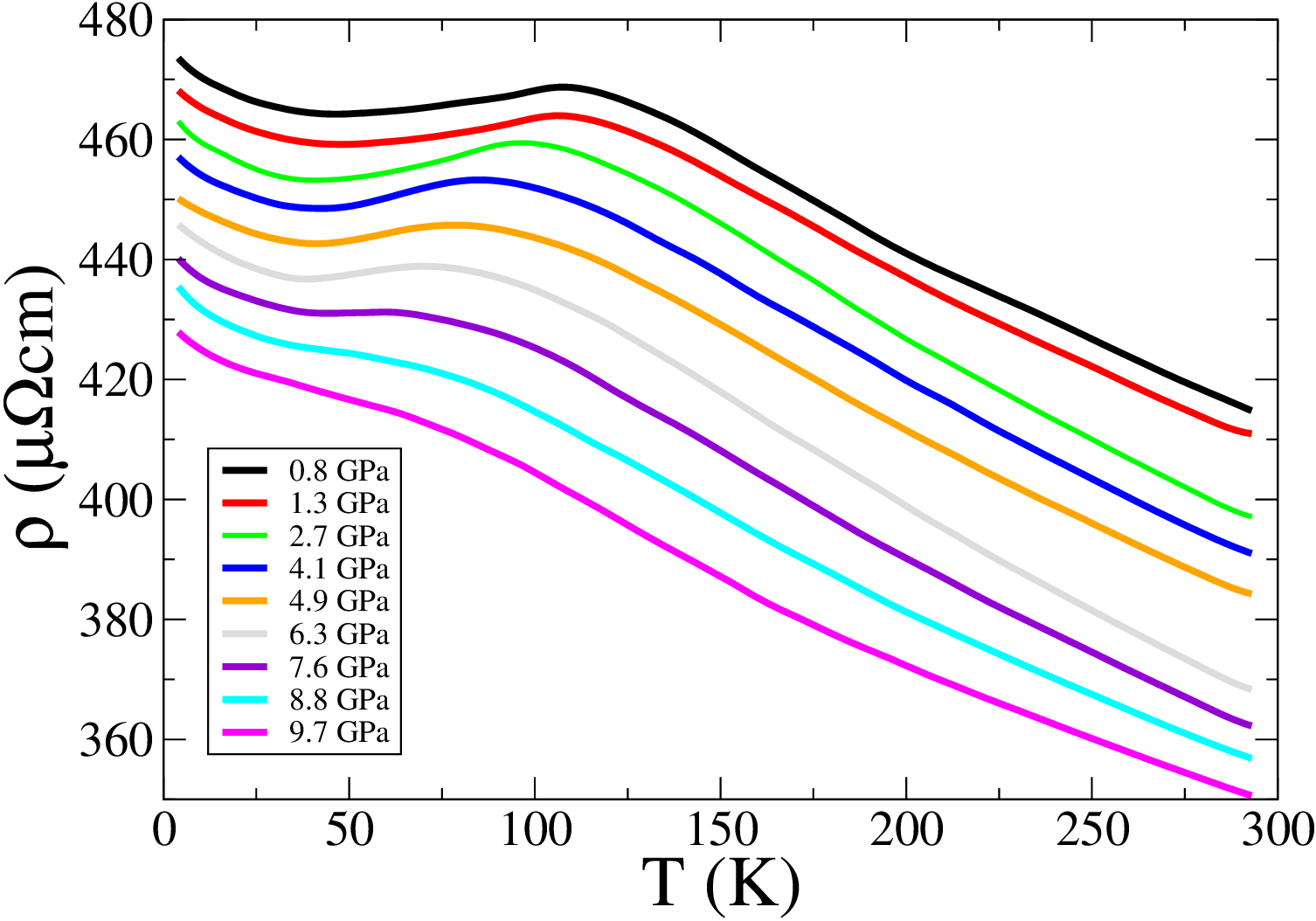}
\caption{ Resistivity of the $c$-axis sample as a function of temperature, 
  from 300 to 2 K, at fixed pressures from 0.8 to 9.7 GPa. 
The peak in each curve marks the antiferromagnetic transition temperature, $T_N$. 
Increasing pressure shifts $T_N$ to lower values, and reduces the resistivity.  
Low temperature data ($T<15$\,K) fit a sub-linear power law with nearly the same 
  exponent ($x=0.7\pm0.1$) at all pressures.
}
\label{fig:rhocvsp}
\end{center}
\end{figure}

Figure \ref{fig:rhocvsp} shows $\rho_c$ vs.\ $T$ at pressures between 0.8 and 
 9.7 GPa. 
The main features of this plot are: 
  1) although the curves shift downwards with increasing pressure, showing that the sample 
     becomes more conducting with increasing pressure, the overall effect is small; 
  2) $T_N$ is suppressed by pressure, as shown by the shift to lower temperature of 
     the peak in $\rho_c$, which is known from our previous measurements to coincide with $T_N$ \cite{wu09};  
     and 
  3) the overall shape of $\rho_c$ vs.\ $T$ does not change markedly. 

Point (3) is our key result.  Despite the fact that $T_N$ is suppressed by more than 
  a factor of two, 
  for $T>T_N$ the resistivity remains non-metallic with little change 
  in slope. 
The $T\rightarrow 0$ K slope is also roughly independent of pressure, remaining 
  non-Fermi-liquid like at all pressures, with the same power law behaviour within the error, 
  $\rho_c \sim \rho_{c,\circ} - A T^{0.7 \pm 0.1}$, as was observed at ambient pressure.

\begin{figure}
\begin{center}
\includegraphics[width=10cm]{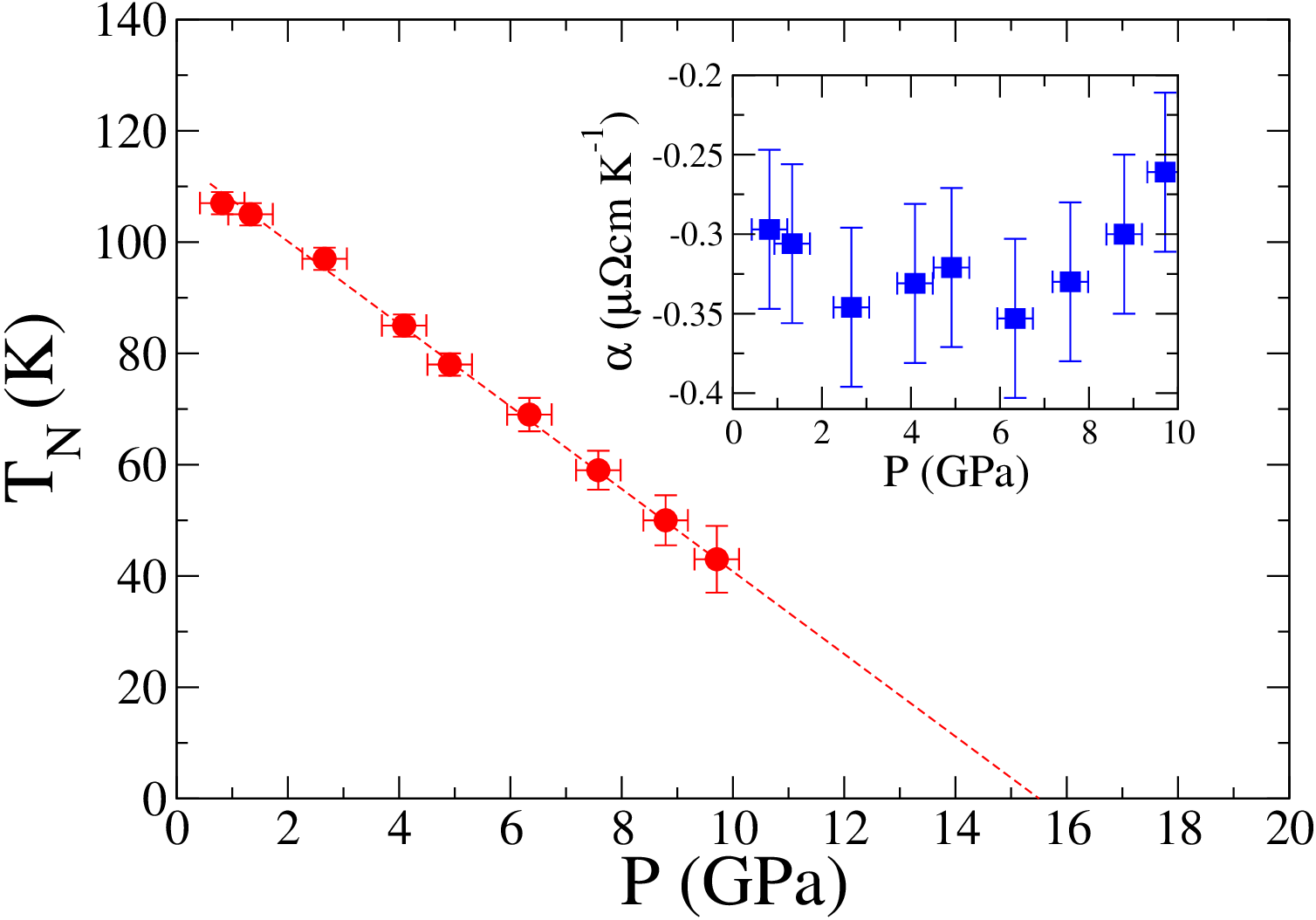}
\caption{ $T_N$, extracted from the data in figure \ref{fig:rhocvsp}, vs.\ pressure. 
A linear extrapolation would place the quantum critical point at $P_c\approx 15.5 \pm 1$\,GPa. 
The inset shows the pressure dependence of the slope $d\rho/dT$ vs.\ 
  $T$ for $T>T_N$, obtained by fitting a linear function to the data between 
  150 and 300 K. 
No significant pressure dependence is observed, showing the robustness of the 
  non-metallic resistivity behaviour against pressure. 
}
\label{fig:TNvsp}
\end{center}
\end{figure}

In figure \ref{fig:TNvsp} we elaborate on points (2) and (3) above. 
The main figure shows a plot of $T_N$ vs.\ $P$. 
$T_N$ has been extracted from the curves in figure \ref{fig:rhocvsp} by finding the 
  maximum in the second derivative.  Note that the maximum weakens with 
  increasing pressure, and ultimately becomes a bump at 9.7 GPa, 
  making it more difficult to extract $T_N$.  
Nevertheless, it is clear from figure 3 that $T_N$ falls roughly linearly with pressure. 
Fitting a straight line to the points in figure 3 gives 
  $dT_N/dP = - 7.3 \pm 0.1 $ K/GPa.  
If we extrapolate this linear behaviour to estimate the pressure at which 
  the quantum critical point, $T_N =0$~K, would be reached, 
  then we find $P_c \sim 15.5 \pm 1$ GPa. 
It should be noted, however, that such extrapolations are not always 
  reliable: $T_N$ vs.\ $P$ curves can turn downwards \cite{mathur98} or 
  saturate \cite{marrec02}, so that $P_c$ may be significantly smaller or larger 
  than this value. 

The inset of figure \ref{fig:TNvsp} shows that the high temperature slope of  
  $\rho_c$ vs.\ $T$ is unaffected by pressure, within the error, 
  emphasizing that pressure has little effect on the non-metallic 
  resistivity at $T>T_N$. 

\begin{figure}
\begin{center}
\includegraphics[width=10cm]{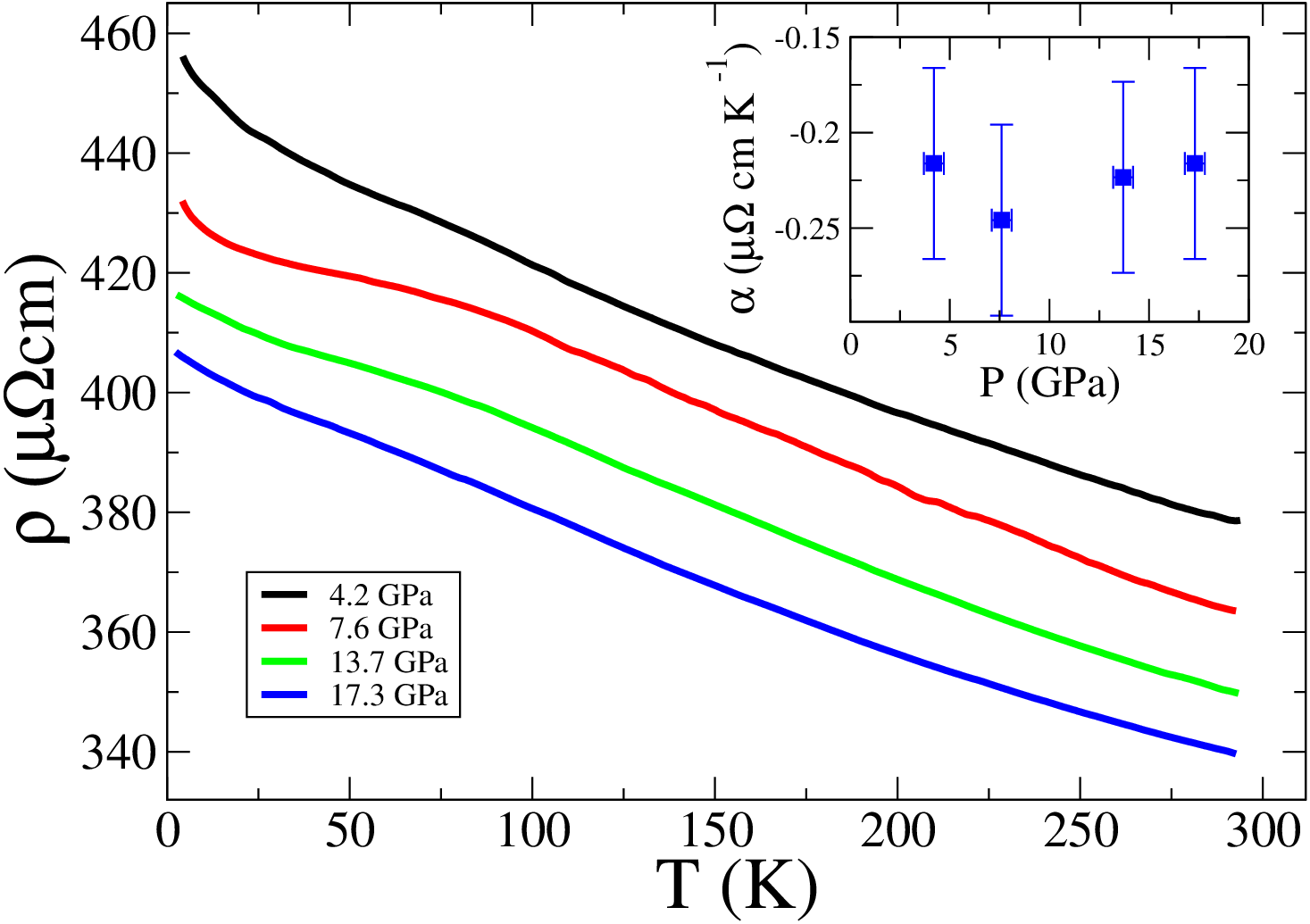}
\caption{ Resistivity of the $\rho_{ab}$ sample as a function of 
  temperature, from 300 to 2 K. 
The low temperature data ($T<15$\,K) fit to a sub-linear power law with 
  nearly the same exponent ($x=0.7\pm0.1$) for the two lowest-pressure 
  curves, but become nearly linear at 13.7 and 17.3 GPa. 
The inset shows the pressure dependence of the $T>T_N$ slope $d\rho/dT$ as a 
  function of pressure, obtained by fitting a line to the 
  data between 150 and 300 K. 
No significant pressure dependence is observed. 
}
\label{fig:rhoabvsp}
\end{center}
\end{figure}

Figure \ref{fig:rhoabvsp} shows $\rho_{ab}$ vs.\ $T$ at four pressures between 
  4.3 and 17.3 GPa. 
Unfortunately, $\rho_{ab}$ does not have a sharp feature 
  at $T_N$ (see figure \ref{fig:ambient}), 
  so we cannot continue our $T_N$ vs.\ $P$ curve using this data.
As with $\rho_c$, $\rho_{ab}$ decreases with increasing 
  pressure, the overall effect is small, 
  and the non-metallic temperature dependence is 
  relatively unaffected by pressure up to 17.3 GPa.  
The high-temperature slope, shown in the inset, is unchanged within the error. 

The $T\rightarrow 0$ K non-metallic behaviour of $\rho_{ab}$ vs.\ $T$ also persists 
  to high pressure, however the slope is smaller in the two highest-pressure curves, 
  and the unusual sub-linear temperature dependence crosses over to become more linear 
  in $T$.  
At intermediate temperatures the concave region,  
  seen at ambient pressure  below $T_N$, seems to have been suppressed in these highest 
  pressure curves.  
This would be consistent with the extrapolation of $T_N$ in figure \ref{fig:TNvsp}, 
  so antiferromagnetism may indeed have been suppressed by 17.3 GPa. 
As a result of the suppression of this concave downwards section, $\rho_{ab}$ at 
  17.3 GPa looks quasilinear over the whole temperature range.  

We have seen no indication of superconductivity in any of our data.  

\section{Discussion:}   \label{sec:discussion}

In these measurements we have suppressed $T_N$ by at least a factor of two, and 
  if the extrapolation of $T_N$ vs.\ $P$ beyond 10 GPa can be trusted, 
  then in our measurements 
  up to 17.3 GPa on the $\rho_{ab}$ sample $T_N$ may have been 
  suppressed to 0 K.
Despite this suppression of antiferromagnetic order, the overall shape of the 
  resistivity curves is not dramatically affected:  the high temperature resistivity 
  remains non-metallic with minimal change of its large, negative slope, 
  while the $T\rightarrow 0$ K resistivity also remains non-metallic, with non-Fermi-liquid 
  power laws persisting to all but the two highest pressures, 
  and with remarkably little change of slope.  
The most notable changes that we do observe are the 
  gradual disappearance of the concave downward regions in $\rho$ vs.\ $T$ that are 
  associated with antiferromagnetic order, and an indication in the $\rho_{ab}$ sample 
  that for $P>10$ GPa the $T\rightarrow 0$ K resistivity 
  crosses over from sublinear to linear dependence on $T$.
This latter change may have to do with  the antiferromagnetic quantum critical 
  point being approached. 
Linear resistivities are typical of 
  antiferromagnetic quantum critical points, although in all of the cases that we know of,  
  the slope of $\rho$ vs.\ $T$ is positive, not negative as in FeCrAs. 

There are few materials to which we can compare these results. 
The robustness of the non-metallic resistivity of FeCrAs under pressure is in sharp 
  contrast to that of CeCuAs$_2$ \cite{sampath05}, 
  whose strongly non-metallic resistivity between room temperature 
  and 1.8 K is completely suppressed at 10 GPa to produce a metallic resistivity 
  over the entire temperature range. 
LaFeAsO is the iron-pnictide superconductor whose resistivity most closely resembles 
  FeCrAs: for $T>200$ K its ambient-pressure resistivity is nearly flat, 
  although unlike FeCrAs its resistivity becomes metallic for $T<200$ K.  
As pressure is increased up to 12 GPa, 
  the magnitude of the high-temperature resistivity falls, but the non-metallic 
  slope remains flat \cite{okada10}.  
Thus, the behaviour is like FeCrAs in that the resistivity 
  vs.\ temperature curves are displaced downwards by pressure but their slope is not 
  changed. 
Another 1111 pnictide, CaFeAsO, has a quasi-linear, weakly metallic, slope at high temperature, 
  and again this slope is nearly independent of pressure while the resistivity curves displace  
  downwards \cite{okada10}. 
The 122 pnictides are even more metallic at high temperature, and in these systems 
  pressure increases the metallic slope somewhat, so pressure makes these materials 
  more metallic, e.g.\ \cite{colombier09}.
One major difference between FeCrAs and all of these systems however is that 
  pressure suppresses the total resistivity much more slowly in FeCrAs. 
For example, in LaFeAsO \cite{okada10}, the non-metallic resistivity at room temperature falls by 
  55\% between 0 and 12 GPa -- from from $3.8$ to $1.7$ m$\Omega$cm -- while 
  in FeCrAs $\rho_{ab}$ at room temperature falls by only 10\% between 4 and 17 GPa 
  -- from $380$ to $340$ $\mu\Omega$cm. 

While our results rule out magnetic long-range order as an important factor in the 
  non-metallic resistivity of FeCrAs at low temperature, they do not necessarily 
  exclude spin-fluctuations as playing a role. 
The iron-pnictides superconductors are believed to be 
  incipient Mott insulators.
In a theoretical study based on this picture, Dai et al.\ decomposed  
  the electronic excitations into a coherent part near the Fermi 
  energy and an incoherent part further away \cite{dai09}.
The latter comprises incipient lower and upper Hubbard bands, 
  which accommodate localized Fe moments.
This model supports a magnetic quantum critical point as a result of the 
  competition between magnetic ordering of the local moments and the mixing of the 
  local moments with the coherent electrons.
The spectral weight of the coherent quasiparticle peak changes as a result of 
  mixing and once it exceeds a critical value, magnetism disappears.
In analogy with this model, the non-metallic behaviour of FeCrAs 
  could be a manifestation of incipient Mott insulating behaviour.

However, there are good reasons for thinking that the physics of FeCrAs may be different. 
In FeCrAs, local moments reside on the Cr and not the Fe sites, 
  so the incoherent carriers would have to be released from Cr $d$-orbitals while the 
  coherent part of the spectrum most likely would stem from Fe $d$-orbitals hybridized with 
  As $p$-orbitals. 
Non-Fermi-liquid behaviour could then arise due to a coupling between the two species of carriers.
Within this theoretical framework, 
  we naively expect that, because pressure increases the mixing between
  the coherent and incoherent parts of the spectrum, 
  at some critical pressure the magnetic order must vanish.
This model has not been worked out for FeCrAs, and as far as we know it 
  would not explain the Fermi liquid specific heat and magnetic susceptibility that are 
  seen at ambient pressure. 

Morever, it is far from clear that FeCrAs is close to a Mott transition.  
As noted in the introduction, 
  band-structure calculations predict three large Fermi surfaces, and would seem 
  to place the system far from a Mott state. 
Alternative models, for example  
  involving orbital effects \cite{haule08}, a hidden spin liquid \cite{rau11}, 
  microscopic phase-separation \cite{dobrosavljevic12}, 
  or even some exotic form of Kondo effect
  may more accurately capture the physics. 

It would be of interest to carry out optical 
  conductivity and NMR measurements to see if a pseudogap is forming as the 
  resistivity rises with decreasing temperature, and to investigate the spin dynamics. 
In terms of possible spin-liquid states \cite{rau11}, thermal conductivity 
  measurements at low temperature may be enlightening.

\section{Conclusions}  \label{sec:conclusions}

We have found that suppression of antiferromagnetic long-range order doesn't 
  strongly affect the non-metallic resistivity seen across several decades of 
  temperature in FeCrAs, while the the non-Fermi-liquid power law behaviour of 
  the resistivity as $T\rightarrow 0$ K, at most crosses over from sublinear 
  below 10 GPa, to linear above 10 GPa.
These results rule out the formation of spin-density-wave 
  gaps on the Fermi surface as playing an important role in the anomalous 
  non-metallic resistivity of this material. 

\section{Acknowledgements} 

We are grateful to Vladimir Dobrosavljevic, Hae-Young Kee and Jeff Rau for helpful discussions. 
This research has been funded by the National Science and Engineering Research 
  Council of Canada and the Canadian Institute for Advanced Research. 

\section*{References}

\providecommand{\newblock}{}

\end{document}